\let\accentvec\vec

\documentclass[runningheads]{llncs}

\let\vec\accentvec

\usepackage{wojtek25preamble,wojtek15logics,wojtek15other}

\newcommand{\Act}{{Act}}
\newcommand{\state}{q}
\renewcommand{\act}{\alpha}
\newcommand{\mov}{\profile{\act}}
\newcommand{\coalition}{C}
\newcommand{\distribution}{\mathsf{d}}
\newcommand{\Dist}{\text{Dist}}
\newcommand{\System}{M}
\newcommand{\legal}{R}
\newcommand{\val}{\Val}
\newcommand{\obsrel}[1][]{\sim_{#1}}
\newcommand{\trace}{\pi}
\renewcommand{\path}{\lambda}
\newcommand{\history}{h}
\newcommand{\History}{\text{Hist}}

\newcommand{\last}{\mbox{last}}
\renewcommand{\str}{\sigma}
\newcommand{\outtr}{\mathit{outtraces}}
\newcommand{\traces}{\mathit{Traces}}
\newcommand{\setstrat}{\mathit{Str}}
\renewcommand{\profile}[1]{\boldsymbol{#1}}
\newcommand{\ag}[1]{^{(#1)}}
\newcommand{\superemph}[1]{{\color{red}\pmb{#1}}}
\newcommand{\cooponly}[1]{\widehat{\coop{#1}}}
\newcommand{\dott}{\, .\,}

\newcommand{\PCTL}[1][]{\lan{PCTL_{\stratstyle{#1}}}}
\newcommand{\PCTLs}[1][]{\lan{PCTL_\stratstyle{#1}^*}}
\newcommand{\PATL}[1][]{\lan{PATL_{\stratstyle{#1}}}}

\newcommand{\PATLs}[1][]{\lan{PATL_\stratstyle{#1}^*}}
\newcommand{\onlyATL}[1][]{\lan{\widehat{ATL_{\stratstyle{#1}}}}}
\newcommand{\onlyATLs}[1][]{\lan{\widehat{ATL_\stratstyle{#1}^*}}}
\newcommand{\compos}     {\mathit{Comp}}

\newcommand{\ie}         {i.e., }

\newcommand{\es}         {\varnothing}
\newcommand{\Nat}        {\mathbb{N}}

\newcommand{\prob}         {\rho}

\renewcommand{\model}{\System}
\newcommand{\assmodel}{\mathcal{A}}

\graphicspath{{./models/}}

\begin{document}

\title{%
Towards Assume-Guarantee Verification of Abilities \\ in Stochastic Multi-Agent Systems} 
\titlerunning{Assume-Guarantee Verification of Abilities in Stochastic MAS} 
\author{
  Wojciech Jamroga\inst{1,2} \and
  Damian Kurpiewski\inst{1,2} \and
  {\L}ukasz Mikulski\inst{2}
}

\institute{
Institute of Computer Science, Polish Academy of Sciences, Warsaw, Poland \and
Faculty of Mathematics and Computer Science, Nicolaus Copernicus University, Toru{\'n}, Poland
}

\maketitle

\begin{abstract}
Model checking of strategic abilities is a notoriously hard problem, even more so in the realistic case of agents with imperfect information, acting in a stochastic environment.
Assume-guarantee reasoning can be of great help here, providing a way to decompose the complex problem into a small set of easier
subproblems.

In this paper, we propose several schemes for assume-guarantee verification of probabilistic alternating-time temporal logic with imperfect information.
We prove the soundness of the schemes, and discuss their completeness. On the way, we also propose a new variant of (non-probabilistic) alternating-time logic, where the strategic modalities capture ``achieving at most $\varphi$,'' analogous to Levesque's logic of ``only knowing''.
\end{abstract}

\keywords{multi-agent systems, formal verification, model checking, strategic ability, assume-guarantee reasoning, probabilistic models and properties}

%
%
%
%

\section{Introduction}\label{sec:intro}

Multi-agent systems involve a complex network of social and technological components.
Such components often exhibit self-interested, goal-directed behavior, which makes it harder to predict and analyze the dynamics of the system.
In consequence, formal specification and automated verification can be of significant help.

\para{Verification of strategic ability.} Many important properties of multi-agent systems refer to \emph{strategic abilities} of agents and their groups.
\emph{Alter\-nating-time temporal logic} \ATLs~\cite{Alur02ATL,Schobbens04ATL} and \emph{Strategy Logic} \SL~\cite{Mogavero14behavioral} provide powerful tools to reason about such aspects of MAS.
For example, the \ATLs formula $\coop{taxi}\Always\neg\prop{fatality}$ expresses that the autonomous cab can drive in such a way that no one gets ever killed. Similarly, $\coop{taxi,passg}\Sometm\prop{destination}$ says that the cab and the passenger have a joint strategy to arrive at the destination, no matter what the other agents do.
Specifications in agent logics can be used as input to algorithms and tools for \emph{model checking},
that have been in constant development for over 20 years~\cite{Alur98mocha-cav,Busard15reasoning,Cermak15mcmas-sl-one-goal,Huang14symbolic-epist,Kurpiewski21stv-demo,Lomuscio17mcmas}.

Model checking of strategic abilities is hard, both theoretically and in practice. First, it suffers from the well-known state/transition-space explosion.
Moreover, the space of possible strategies is at least exponential \emph{on top of the state-space explosion}, and incremental synthesis of strategies is not possible in general -- especially in the realistic case of agents with partial observability.
Even for the more restricted (and computation-friendly) logic \ATL, model checking of its imperfect information variants is \Deltwo- to \Pspace-complete for agents playing memoryless strategies~\cite{Bulling10verification,Schobbens04ATL} and \EXPTIME-complete to undecidable for agents with perfect recall~\cite{Dima11undecidable,Guelev11atl-distrknowldge}.
The theoretical results concur with outcomes of empirical studies on benchmarks~\cite{Busard15reasoning,Jamroga19fixpApprox-aij,Lomuscio17mcmas}, as well as recent attempts at verification of real-life multi-agent scenarios~\cite{Jamroga20Pret-Uppaal,Kurpiewski19embedded}.

The idea of \emph{assume-guarantee reasoning}~\cite{Clarke89assGuar,Pnueli84assGuar} is to ``factorize'' the verification task into subtasks where components are verified against a suitable abstraction of the rest of the system.
Thus, instead of searching through the states (and, in our case, strategies) of the huge product of all components, most of the search is performed locally.

\para{Contribution.}
In this paper, we extend our previous work on assume-guarantee verification~\cite{Mikulski22towardsAGV,Mikulski22AGV,Kurpiewski22STV+AGV} to the analysis of strategic ability in stochastic multi-agent systems.
Formally, we propose several assume-guarantee schemes for \PATL (i.e., Probabilistic \ATL~\cite{Chen07PATL}), some of them simpler and easier to automatize, the other ones more sophisticated but complete. This way, we make the first step towards compositional model checking of strategic properties in stochastic multi-agent systems with imperfect information, which is a distinctly hard problem~\cite{Belardinelli23PATL,Belardinelli24PATL,Jamroga25PTATL}.

Moreover, to formalize some of the schemes, we propose a new variant of alternating-time temporal logic, \onlyATLs, that captures the ability of agents to enforce \emph{no more than $\varphi$} (similar to the ``only knowing'' or ``knowing no more that'' modalities of Levesque, Halpern, and Lakemeyer~\cite{Levesque90alliknow,Halpern01onlyknowing}). We also study some metaproperties of the logic -- in particular, we show that it has the same model checking complexity as standard \ATL, but allows for expressing properties that are not definable in standard \ATL.


We emphasize that our schemes are sound for the model checking of agents with \emph{deterministic strategies} and \emph{imperfect} as well as \emph{perfect recall}. In consequence, they can be used to facilitate verification problems with a high degree of hardness, including the undecidable variant for coalitions of agents with memory.

\para{Structure of the paper.}
We start by discussing the related work in Section~\ref{sec:related}. Then, we recall the logical frameworks of \ATLs and \PATLs, as well as the assume-guarantee framework of~\cite{Mikulski22AGV} for strategic abilities in non-probabilistic models (Sections~\ref{sec:atl+agv} and~\ref{sec:patl}).
We discuss the problems with extending the framework to strategies in probabilistic MAS, propose the general idea to overcome the problems, and formalize it in the form of $3$ alternative assume-guarantee schemes (Sections~\ref{sec:towards-idea} and~\ref{sec:agv-patl}).
Section~\ref{sec:agv-patl-onlyachieving} is centered around a new variant of \ATL for ``achieving no more than $\varphi$.'' We define its syntax and semantics, establish the expressivity and model checking complexity, and finally use it to propose two more schemes of assume-guarantee reasoning for strategic abilities in stochastic environments.
We conclude in Section~\ref{sec:conclusions}.

\section{Related Work}\label{sec:related}

Compositional verification (known as \emph{rely-guarantee} in the program verification community) dates back to the early 1970s and the works of Hoare, Owicki, Gries and Jones~\cite{Hoare69axiomatic,Jones83relyGuar,Owicki76relyGuar}.
Assume-guarantee reasoning for temporal specifications was introduced a decade later~\cite{Clarke89assGuar,Pnueli84assGuar}, and has been in development since that time~\cite{Devereux03compositionalRA,Fijalkow20assGuar,Henzinger98assGuar,Kwiatkowska10assGuar,Lomuscio10assGar,Lomuscio13assGar}. Moreover, automated synthesis of assumptions for temporal reasoning has been studied in~\cite{Chen10assGuar-learning,Giannakopoulou05assGuar-learning,He16assGuar-learning,Kong10assGuar-learning}.

In~\cite{Lomuscio10assGar,Lomuscio13assGar}, models and a reasoning scheme were defined for assume-guarantee verification of liveness properties in distributed systems. The approach was extended to the verification of strategic abilities in~\cite{Mikulski22towardsAGV,Mikulski22AGV,Kurpiewski22STV+AGV}. We build directly on those semantic frameworks, and extend their solutions to strategic reasoning for agents acting in a stochastic environment.
Other works concerning assume-guarantee verification of strategic ability include~\cite{Devereux03compositionalRA} where assume-guarantee reasoning for an early version of \ATL was studied. We not in passing that their assume-guarantee rules were designed for perfect information strategies (whereas we tackle the more complex case of imperfect information), and targeted specifically the verification of aspect-oriented programs.
In~\cite{Majumdar20AGV-synthesis}, an assume-guarantee scheme for synthesis of reactive modules satisfying a given \LTL objective was proposed.
Furthermore, \cite{Finkbeiner22asv-ATL}~investigated the compositional synthesis of finite-memory strategies for \LTL objectives. The advantage of their solution is the use of contracts, thanks to which it is possible to synthesize individual strategies using the knowledge of the coalition partners' strategies. We do not explore this idea here, leaving it for possible future work.
We also mention~\cite{Chatterjee07assume-guarantee} that studied the synthesis of Nash equilibrium strategies for 2-player coalitions pursuing $\omega$-regular objectives. The authors called their approach \emph{assume-guarantee strategy synthesis}, but the connection to assume-guarantee verification is rather loose.


Assume-guarantee reasoning for probabilistic systems and properties has been studied in~\cite{Delahaye08probAG-contracts} for reasoning about trace properties of programs, and then in~\cite{Kwiatkowska10assGuar} with respect to safety properties of probabilistic automata networks.
\cite{Komuravelli12AGV+AR-prob}~proposed assume-guarantee abstraction refinement for Labeled Probabilistic Transition Systems.
In a related line of research~\cite{Chatterjee14CEGAR+AGV-prob}, counterexample-guided abstraction refinement was used to provide automated assume-guarantee verification for the qualitative fragment of \PCTL (i.e, the fragment of Probabilistic \CTL that only asks of achieving the objective with probability $1$).
More recently, \cite{Zimmermann25robustPCTL}~proposed robust versions of \PCTL and \PCTLs, that can embed assume-guarantee reasoning in their object language.
Importantly, none of the above works considered assume-guarantee reasoning about probabilistic strategic abilities, neither individual, nor coalitional.

The verification schemes, proposed and studied here, are an extension of those in~\cite{Mikulski22AGV} to reasoning about abilities in probabilistic multi-agent systems. As it turned out, we needed to extend the scheme of~\cite{Mikulski22AGV} in a non-trivial way, including a definition of a completely new strategic logic of ``achieving at most,'' rather than -- as it is usually the case -- ``achieving at least'' a given path property $\varphi$.

\section{Modelling and Verification of Agent Interaction}\label{sec:atl+agv}

We begin by summarizing the main formal concepts of~\cite{Mikulski22AGV} and~\cite{Chen07PATL,Jamroga25PTATL}, in particular formal models of concurrent multi-agent systems, syntax and semantics of \ATL and \PATL, and the assume-guarantee schemes proposed in~\cite{Mikulski22AGV} for model checking of \ATL.

\subsection{Models of Concurrent MAS}\label{sec:models}

Asynchronous MAS have been modeled by variants of reactive modules~\cite{Alur99reactive,Lomuscio13assGar} and automata networks~\cite{Jamroga21paradoxes-kr}.
Here, we adapt the variant of reactive modules that was used to define assume-guarantee verification for temporal properties in~\cite{Lomuscio13assGar}.

\subsubsection{Modules.}\label{sec:modules}
Let $D$ be the shared domain of values for all the variables in the system.
$D^X$ is the set of all valuations for a set of variables $X$.
The \emph{system} consists of a number of \emph{agents}, each represented by its \emph{module} and a \emph{repertoire} of available choices.
Every agent uses \emph{state variables} and \emph{input variables}.
It can read and modify its state variables at any moment, and their valuation is determined by the current state of the agent.
The input variables are not a part of the state, but their values influence
transitions that can be executed.

\begin{definition}[Module~\cite{Lomuscio13assGar}]\label{d:module}
A \emph{module} is a tuple $M=(X,I,Q,T,\lambda,q_0)$, where:\
 $X$ is a finite set of state variables;
 $I$ is a finite set of input variables with $X\cap I=\es$;
 $Q=\{q_0,q_1,\ldots,q_n\}$ is a finite set of states;
 $q_0\in Q$ is an initial state;
 $\lambda:Q\to D^X$ labels each state with a valuation of the state variables;
 finally, $T\subseteq Q\times D^I\times Q$ is a transition relation such that
  (a) for each pair $(q,\alpha)\in Q\times D^I$ there exists $q'\in Q$ with $(q,\alpha,q')\in T$, and
  (b) $(q,\alpha,q')\in T, q\neq q'$ implies $(q,\alpha,q)\notin T$.
In what follows, we omit the self-loops from the presentation.
\end{definition}

Modules $M,M'$ are \emph{asynchronous} if $X\cap X'=\es$.
We extend modules by adding \emph{repertoire functions} that define the agents' available choices in a way similar to~\cite{Jamroga21paradoxes-kr}.

\begin{definition}[Repertoire]
Let $M=(X,I,Q,T,\lambda,q_0)$ be a module of agent $i$.
The \emph{repertoire} of $i$ is defined as $R:Q\to \powerset{\powerset{T}}$, i.e., a mapping from local states to \emph{sets of sets of transitions}.
Each $R(q) = \{T_1,\dots,T_m\}$ must be nonempty and consist of nonempty sets $T_i$ of transitions starting in $q$.
If the agent chooses $T_i \in R(q)$, then only a transition in $T_i$ can be occur at $q$ within the module.
\end{definition}

\subsubsection{Composition of Agents.}\label{sec:composition}\label{sec:models-properties}
On the level of the temporal structure, the model of a multi-agent system is given by the asynchronous composition $M = M\ag{1} | \dots | M\ag{n}$ that combines modules $M\ag{i}$ into a single module.
The definition is almost the same as in~\cite{Lomuscio13assGar}; we only extend it to handle the repertoire functions that are needed to characterize strategies and strategic abilities.

We begin with the notion of compatible valuations to adjust local states of one agent
with the labels of the actions performed by the other agent.
Note that the local states of different asynchronous agents rely on disjoint sets of variables.

Let $Y,Z\subseteq X$ and $\rho_1\in D^Y$ while $\rho_2\in D^Z$.
We say that $\rho_1$ is compatible with $\rho_2$ (denoted by $\rho_1 \sim \rho_2$)
if for any $x\in Y\cap Z$ we have $\rho_1(x)=\rho_2(x)$.
We can compute the union of $\rho_1$ with $\rho_2$ which is compatible with $\rho_1$ by setting
$(\rho_1 \cup \rho_2)(x) = \rho_1(x)$ for $x\in Y$ and
$(\rho_1 \cup \rho_2)(x) = \rho_2(x)$ for $x\in Z$.

\begin{definition}[Composition of modules~\cite{Lomuscio13assGar}]\label{d:comp}
The composition of asynchronous modules
$M\ag{1}=(X\ag{1},I\ag{1},Q\ag{1},T\ag{1},\lambda\ag{1},q\ag{1}_{0})$ and \linebreak
$M\ag{2}=(X\ag{2},I\ag{2},Q\ag{2},T\ag{2},\lambda\ag{2},q\ag{2}_{0})$
(with $X\ag{1}\cap X\ag{2}=\es$) is
a composite module
$M=(X=X\ag{1}\uplus X\ag{2},I=(I\ag{1}\cup I\ag{2})\setminus X,Q\ag{1}\times Q\ag{2},T,\lambda,q_0=(q\ag{1}_{0},q\ag{2}_{0}))$,
where
\begin{itemize}
\item $\lambda:Q\ag{1}\times Q\ag{2}\to D^X$, $\lambda(q\ag{1},q\ag{2})=\lambda\ag{1}(q\ag{1})\cup\lambda\ag{2}(q\ag{2})$,
\item $T$ is the minimal transition relation derived by the set of rules presented below:
\[
\bf{ASYN_L}\;\;\;
\begin{array}{c}
q\ag{1}\xrightarrow[]{\alpha\ag{1}}_{T\ag{1}}{q'}\ag{1}\;\;\;\;q\ag{2}\xrightarrow[]{\alpha\ag{2}}_{T\ag{2}}{q'}\ag{2}\\
\alpha\ag{1}\sim\alpha\ag{2} \;\;\;\; \lambda\ag{1}(q\ag{1})\sim\alpha\ag{2} \;\;\;\; \lambda\ag{2}(q\ag{2})\sim\alpha\ag{1}\\
\hline
(q\ag{1},q\ag{2})\xrightarrow[]{(\alpha\ag{1}\cup\alpha\ag{2})\setminus X}_T({q'}\ag{1},q\ag{2})
\end{array}
\]
\[
\bf{ASYN_R}\;\;\;
\begin{array}{c}
q\ag{1}\xrightarrow[]{\alpha\ag{1}}_{T\ag{1}}{q'}\ag{1}\;\;\;\;q\ag{2}\xrightarrow[]{\alpha\ag{2}}_{T\ag{2}}{q'}\ag{2}\\
\alpha\ag{1}\sim\alpha\ag{2} \;\;\;\; \lambda\ag{1}(q\ag{1})\sim\alpha\ag{2} \;\;\;\; \lambda\ag{2}(q\ag{2})\sim\alpha\ag{1}\\
\hline
(q\ag{1},q\ag{2})\xrightarrow[]{(\alpha\ag{1}\cup\alpha\ag{2})\setminus X}_T(q\ag{1},{q'}\ag{2})
\end{array}
\]
\[
\;\;\;\;\;\bf{SYN}\;\;\;
\begin{array}{c}
q\ag{1}\xrightarrow[]{\alpha\ag{1}}_{T\ag{1}}{q'}\ag{1}\;\;\;\;q\ag{2}\xrightarrow[]{\alpha\ag{2}}_{T\ag{2}}{q'}\ag{2}\\
\alpha\ag{1}\sim\alpha\ag{2} \;\;\;\; \lambda\ag{1}(q\ag{1})\sim\alpha\ag{2} \;\;\;\; \lambda\ag{2}(q\ag{2})\sim\alpha\ag{1}\\
\hline
(q\ag{1},q\ag{2})\xrightarrow[]{(\alpha\ag{1}\cup\alpha\ag{2})\setminus X}_T({q'}\ag{1},{q'}\ag{2})
\end{array}
\]
\end{itemize}
pruned in order to avoid disallowed self-loops.
We use the notation $M=M\ag{1}|M\ag{2}$.
\end{definition}

Note that the operation is defined in~\cite{Lomuscio13assGar} for a pair of modules only.
It can be easily extended to a larger number of pairwise asynchronous modules. Moreover, the order of the composition does not matter.

Consider agents $(M\ag{1},R\ag{1}),\ \dots,\ (M\ag{n},R\ag{n})$.
The \emph{multi-agent system} is defined by $\model = (M\ag{1}|M\ag{2}|\ldots|M\ag{n},\ R\ag{1},\ldots, R\ag{n})$, i.e., the composition of the underlying modules, together with the agents' repertoires of choices.

%

\subsubsection{Traces and Words. }\label{sec:traces}
A trace of a module $M$ is an infinite sequence of alternating states and transitions $\trace=q_0\alpha_0 q_1\alpha_1\ldots$, where
$(q_i,\alpha_i,q_{i+1})\in T$ for every $i\in\Nat$ (note that $q_0$ is the initial state).
An infinite word $w=v_0 v_1,\ldots \in(D^X)^\omega$ is \emph{derived} by $M$ with trace $\trace=q_0\alpha_0 q_1\alpha_1\ldots$ if $v_i = \lambda(q_i)$ for all $i\in\Nat$.
An infinite word $u=\alpha_0 \alpha_1,\ldots \in(D^I)^\omega$ is \emph{admitted} by $M$ with $\trace$ if $\trace=q_0\alpha_0 q_1\alpha_1\ldots$.
Finally, $w$ (resp.~$u$) is derived (resp.~admitted) by $M$ if there exists a trace of $M$ that derives (resp.~admits) it.

\subsection{Alternating-Time Temporal Logic}\label{sec:logic}

\emph{Alternating-time temporal logic} \ATLs~\cite{Alur97ATL,Alur02ATL,Schobbens04ATL}
introduces \emph{strategic modalities} $\coop{\coalition}\psi$, expressing that coalition $\coalition$ can enforce the temporal property $\psi$.
We use the semantics based on \emph{imperfect information strategies} with \emph{imperfect recall} ($\ir$) or \emph{perfect recall} ($\iR$)~\cite{Schobbens04ATL}.
Moreover, we only consider formulas without the next step operator $\Next$ due to its questionable interpretation for asynchronous systems, which are based on the notion of local clocks.

\para{Syntax.}
Formally, the syntax of $\ATLsX$ is as follows:
\begin{displaymath}
\phi ::= p(Y) \mid \neg\phi \mid \phi\land\phi \mid \coop{\coalition} \psi\, , \qquad\qquad
\psi ::= \phi \mid \neg\psi \mid \psi\land\psi \mid \psi \Until \psi
\end{displaymath}
where $p:Y\to D$ for some subset of domain variables $Y\subseteq X$. That is, each atomic statement refers to the valuation of variables used in the system.
$\Until$ is the ``strong until'' operator of \LTLX. The ``sometime'' and ``always'' operators can be defined as usual by $\Sometm\psi \equiv \top \Until \psi$ and $\Always\psi \equiv \neg\Sometm\neg\psi$.
The set of variables used by the formula $\psi$ is denoted by $var(\psi)$.

In most of the paper, we focus on formulas that consist of a single strategic modality followed by an \LTLX formula (i.e., $\coop{\coalition}\psi$, where $\psi\in\LTLX$).
The corresponding fragment of \ATLsX, called \oneATLsX, suffices to express many interesting specifications, namely the ones that refer to agents' ability of enforcing trace properties (such as safety or reachability of a winning state).
Note that \oneATLsX has strictly higher expressive and distinguishing power than \LTLX. In fact, model checking \oneATLsX is equivalent to \LTLX controller synthesis, i.e., a variant of \LTL realizability.

Nested strategic modalities might be sometimes needed to refer to an agent’s ability to endow or deprive another agent with/of ability.
We discuss assume-guarantee verification for such specifications in Section~\ref{sec:nested}.

\para{Strategies and Their Outcomes.} 
Let $\model$ be a system composed of $n$ agents with asynchronous modules $M\ag{i} = (X\ag{i},I\ag{i},Q\ag{i},T\ag{i},\lambda\ag{i},q_{0}\ag{i})$ and repertoires $R\ag{i}$.

\begin{definition}[Strategies]
A \emph{memoryless strategy} for agent $i$ (\ir-strategy in short) is a function
$s_i^{\ir}:Q\ag{i}\to \powerset{\powerset{T\ag{i}}}$ such that $s_i^{\ir}(q\ag{i})\in R\ag{i}(q\ag{i})$ for every $q\ag{i}\in Q\ag{i}$.
That is, a memoryless strategy assigns a legitimate choice to each local state of $i$.

A \emph{perfect recall strategy} for $i$ (\iR-strategy in short) is a function $s_i^{\iR}:(Q\ag{i})^+\to T\ag{i}$ such that $s_i^{\iR}(q\ag{i}_1,\dots,q\ag{i}_k)\in R\ag{i}(q\ag{i}_k)$, i.e., it assigns choices to finite sequences of local states.
We assume that $s_i^{\iR}$ is stuttering-invariant, i.e.,
$$s_i^{\iR}(q\ag{i}_1,\dots,q\ag{i}_j,q\ag{i}_j,\dots,q\ag{i}_k) = s_i^{\iR}(q\ag{i}_1,\dots,q\ag{i}_j,\dots,q\ag{i}_k).$$
Note that the agent's choices in a strategy depend only on its \emph{local} states, thus being uniform by construction.
\end{definition}

Let $\trace=q_0\alpha_0 q_1\alpha_1\ldots$ be a trace, where $q_j=(q_j\ag{1},q_j\ag{2},\ldots,q_j\ag{n})$ are global states in $Q\ag{1}\times\ldots\times Q\ag{n}$.
We say that $\trace$ \emph{implements} strategy $s_i^{\ir}$
if, for any $j$ where $q_j\ag{i}\neq q_{j+1}\ag{i}$, we have $(q_j\ag{i},\alpha_j,q_{j+1}\ag{i})\in s_i^{\ir}(q_j\ag{i})$
where $\alpha_j:I\ag{i}\to D$ and $\alpha_j(x)=\lambda(q_j)(x)$.
A word $w=v_0v_1\dots$ \emph{implements} $s_i^{\ir}$
if it is derived by $\model$ with some trace $\trace$ implementing $s_i^{\ir}$.
The definitions for $s_i^{\iR}$ are analogous.

\begin{definition}[Coalitional strategies]
Let $\coalition\subseteq \{1,\ldots,n\}$ be a coalition of agents.
A \emph{joint memoryless strategy $s_\coalition^{\ir}$} for $\coalition$ is a collection of memoryless strategies $s_i^\ir$, one per $i\in \coalition$.
We say that a trace $\trace$ (respectively a word $w_\trace$) \emph{implements} $s_\coalition^{\ir}$ if it implements every strategy $s_i^{\ir}, i\in \coalition$.
The definitions for joint perfect recall strategies are analogous.
Whenever a claim holds for both types of strategies, we will refer to them simply as ``strategies.''
\end{definition}

\para{Semantics.} 
Let $x\in\set{\ir,\iR}$ be a strategy type. The semantics of \ATLsX is given below (we omit the standard clauses for Boolean operators etc.). By $w[i]$, we denote the $i$th item of sequence $w$, starting from $0$.

\begin{description}
\item[{$\model,q \satisf[x] p(Y)$}] if $\lambda(q)|_Y = p(Y)$;

\item[{$\model,q \satisf[x] \coop{\coalition} \psi$}] if there exists an $x$-strategy $s_\coalition$ for $\coalition$ such that, for any word $w$ starting in $q$ that implements $s_\coalition$, we have $\model,w \satisf \psi$;

\item[{$\model,w \satisf \phi$}] if $\model,w[0] \models \phi$;

\item[{$\model,w \satisf \psi_1 \Until \psi_2$}] if there exists $j$ such that $\model,w[j,\infty] \satisf \psi_2$, and $\model,w[i,\infty] \satisf \psi_1$ for each $0\le i<j$.
\end{description}

Finally, we say that $\model \satisf[x] \phi$ if $\model,q_0 \satisf[x] \phi$, where $q_0$ is the initial state of $\model$.

%

\subsection{Assumptions and Guarantees}\label{sec:assumptions}

Our assume-guarantee scheme reduces the complexity of model checking by ``factorizing'' the task into verification of strategies of single agents with respect to abstractions of the rest of the system.
In this section, we formalize the notions of \emph{assumption} and \emph{guarantee}, which provide the abstractions in a way that allows for simulating the global behavior of the system.

\subsubsection{Assumptions.}\label{sec:ass}
For assumptions, we use B\"{u}chi accepting conditions.
More precisely, the infinite word $w=q_0 q_1,\ldots$ is \emph{accepted}
by extended module $(M,F)$ with computation $u=\alpha_0\alpha_1\ldots$
if it is derived by $M$ with a trace $\trace=q_0\alpha_0 q_1\alpha_1\ldots$ and
$\mathit{inf}(\trace)\cap F\neq\es$.
Thus, the assumptions have the expressive power of $\omega$-regular languages.
In practical applications, it might be convenient to formulate actual assumptions in \LTL (which covers a proper subclass of $\omega$-regular properties).

\begin{definition}[Assumption~\cite{Lomuscio13assGar}]
An \emph{assumption} or an \emph{extended module} $(M,F)=(X,I,Q,T,\lambda,q_0,F)$ is a module augmented with a set of accepting states $F\subseteq Q$.
\end{definition}

The definitions of Sections~\ref{sec:models} and~\ref{sec:logic} generalize to assumptions in a straightforward way.
In particular, we can compose a module $M$ with an assumption $A'=(M',F')$, and obtain an extended composite module $A=(M|M',F)$,
where $F = \{(q,q')\in Q\times Q' \mid q'\in F'\}$.
We use the notation $A=M|A'$.
Moreover, let $\assmodel = (A,R\ag{1},\dots,R\ag{m})$ be a MAS based on the extended module $A$
with repertoires related to all components of $M$.
The semantics of \oneATLsX extends naturally:

\begin{description}
\item[{$\assmodel,q \satisf[x] \coop{\coalition} \psi$}] iff there exists an $x$-strategy $s_\coalition$ for $\coalition$ such that,
for any word $w=w[1]w[2]\ldots$ that implements $s_\coalition$ and is accepted by $A$, we have $\assmodel,w \satisf[x] \psi$.
\end{description}

%
%

\subsubsection{Guarantees.}\label{sec:guar}
We say that a sequence $v=v_1v_2\ldots$ over $D^Y$ is
a \emph{curtailment} of a sequence $u=u_1u_2\ldots$ over $D^X$ (where $Y\subseteq X$)
if there exists an infinite sequence $c$ of indices $c_0<c_1<...$ with $c_0=0$
such that $\forall_i \forall_{c_i\leq k<c_{i+1}} v_i=u_k|_Y$.
We will denote a curtailment of $u$ to $D^Y$ by $u|_Y$ or $u|_Y^c$,
and use it to abstract away from irrelevant variables and the stuttering of states.

\begin{definition}[Guarantee~\cite{Lomuscio13assGar,Mikulski22AGV}]
Let $M\ag{1}, \ldots, M\ag{k}$ be pairwise asynchronous modules, and
$A=(X\ag{A},I\ag{A},Q\ag{A},T\ag{A},\lambda\ag{A},q\ag{A}_0,F\ag{A})$ be an assumption
with $X\ag{A}\subseteq X=\bigcup_{i=1}^k X\ag{i}$ and $I\ag{A}\subseteq I=\bigcup_{i=1}^k I\ag{i}$.

We say that $M=M\ag{1}|\ldots|M\ag{k}$ \emph{guarantees} the assumption $A$ (denoted $M\models A$)
if, for every infinite trace $\trace$ of $M$ with $w\in (D^X)^\omega$ derived by $M$ with $\trace$ and $u\in (D^I)^\omega$ admitted by $M$ with $\trace$,
there exists a curtailment $w|_{X\ag{A}}^c$ ($c=c_1,c_2,\ldots$) accepted by $A$ with the computation $u_{c_1-1}|_{I\ag{A}}\;u_{c_2-1}|_{I\ag{A}}\;\dots$ .
\end{definition}
That is, every trace of $M$ must agree on the values of $X\ag{A}$ with some trace in $A$, modulo stuttering.

%
%

It is possible to relate the traces of a subsystem with the traces of the entire system in such a way
that it is possible to verify locally defined formulas.

\subsection{\mbox{Assume-Guarantee Reasoning for 1ATL*}}\label{sec:agv-single-agents}

Now we recall the assume-guarantee schemes of~\cite{Mikulski22AGV} that decompose abilities of coalition $\coalition$ into abilities of its subcoalitions, verified in suitable abstractions of their neighbor modules.

\subsubsection{Assume-Guarantee Rule for Strategies.}\label{sec:agv-individual}
Let $\model$ be a system composed of asynchronous agents $(M\ag{1},R\ag{1}),\ \dots,\ (M\ag{n},R\ag{n})$.
By $N\ag{i}_1$, we denote the direct ``neighborhood'' of agent $i$, i.e., the set of agent indices $j$ such that $I_{M\ag{j}}\cap X_{M\ag{i}}\neq\es$ or $I_{M\ag{i}}\cap X_{M\ag{j}}\neq\es$.
By $N\ag{i}_k$, we denote the agents connected to $i$ in at most $k$ steps, i.e., $(N\ag{i}_{k-1}\cup\bigcup_{j\in N\ag{i}_{k-1}}N\ag{j}_1) \setminus \{i\}$.
Finally $\compos\ag{i}_k$ denotes the composition of all modules of $N\ag{i}_k$.
That is, if $N\ag{i}_k=\{a_1,...,a_m\}$ then $\compos\ag{i}_k=M\ag{a_1} | ... | M\ag{a_m}$.

Let $\psi_{i}$ be an \LTL formula (without ``next''), where atomic propositions are local valuations of variables in $M\ag{i}$.
Also, let $x\in\{\ir,\iR\}$.
The scheme is formalized through a sequence of rules $\bf{R_k}$ which rely on the behaviour of the neighbourhoods of coalition $\coalition$, limited by ``distance'' $k$:

\[
\bf{R_k}\;\;\;
\begin{array}{c}
\forall_{i\in \coalition}\; (M\ag{i} | A_i, R\ag{i}) \,\satisf[x]\, \coop{i} \psi_{i}\\
\forall_{i\in \coalition}\; \compos\ag{i}_k \,\models\, A_i\\
\hline
(M\ag{1} | ... | M\ag{n},R\ag{1},\dots,R\ag{n}) \,\satisf[x]\, \coop{\coalition} \bigwedge_{i\in \coalition}\psi_{i}
\end{array}
\]

The main challenge in applying the scheme is to define the right assumptions and to decompose the verified formula.

The following theorem says that, if each coalition member together with its assumption satisfies the decomposition of the formula, and its neighborhood satisfies the assumption, then the original verification task must return ``true.''

\begin{theorem}[\cite{Mikulski22AGV}]\label{t:rksound}
The rule $\bf{R_k}$ is sound.
\end{theorem}

Unfortunately, there does not always exist $k<n$ for which the rule $\bf{R_k}$ is complete,
even in a very weak sense, where we only postulate the \emph{existence} of appropriate assumptions.

\begin{theorem}[\cite{Mikulski22AGV}]
The scheme consisting of rules $\set{\bf{R_k} \mid k\in\Nat}$ is in general not complete.
\end{theorem}

\subsubsection{Coalitional Assume-Guarantee Verification}\label{sec:agv-coalitions}

A possible way out is to allow for assume-guarantee reasoning about joint strategies of subcoalitions of $\coalition$.
This can be achieved by partitioning the system into smaller subsystems and allowing to explicitly consider the cooperation between coalition members.

Again, let $\model=(M\ag{1},R\ag{1}),\ \dots,\ (M\ag{n},R\ag{n})$ be a system composed of asynchronous agents .
Moreover, let $\{P_1, \ldots, P_k : P_i\subseteq\{1,2,\ldots,n\}\}$,
be a partitioning of coalition $\coalition$,
and let $\overline{\coalition}=\{i : i\notin \coalition\}=Ag\setminus \coalition$ be the set of opponents of $\coalition$.
By $\model\ag{P_i}$ we denote the system composed of all the agents in $P_i = \{i_1,\ldots,i_s\}$, i.e., $(M\ag{P_i}=M\ag{i_1}|\ldots|M\ag{i_s}, R\ag{i_1},\ldots,R\ag{i_s})$.
$\model\ag{\overline{\coalition}}$ is defined analogously.

We extend the notion of neighbourhood to sets of agents as follows:\
\begin{itemize}
\item $N^{P_i}_1=(\bigcup_{i\in P_i}N\ag{i}_1)\setminus P_i$,\
$N^{P_i}_k=(N^{P_i}_{k-1}\cup\bigcup_{j\in N^{P_i}_{k-1}}N\ag{j}_1) \setminus P_i$ for $k>1$,\
\item $\compos^{P_i}_k=M\ag{x_1} | ... | M\ag{x_s}$ for $N^{P_i}_k=\{x_1,...,x_s\}$.
\end{itemize}

Let $x\in \{\ir,\iR\}$. The generalized assume-guarantee rule is defined below:

\[
\bf{Part^P_k}\;\;\;
\begin{array}{c}
\forall_{P_i\in P}\; (M\ag{P_i} | A_i,R\ag{i_1},\ldots,R\ag{i_s}) \,\models_{x}\, \coop{P_i} \bigwedge_{j\in P_i}\psi_j\\
\forall_{P_i\in P}\; \compos^{P_i}_k \,\models\, A_i\\
\hline
(M\ag{1} | ... | M\ag{n},R\ag{1},\ldots,R\ag{n}) \models_{x} \coop{\coalition} \bigwedge_{i\in \coalition}\psi_i
\end{array}
\]

As it turns out, the new scheme is sound and complete.

\begin{theorem}[\cite{Mikulski22AGV}]\label{t:partsound}
The rule $\bf{Part^P_k}$ is sound.
\end{theorem}

\begin{theorem}[\cite{Mikulski22AGV}]\label{t:complete}
There exist a partition set $P$ and $k\leq n$ such that the rule $\bf{Part^P_k}$ is complete.
\end{theorem}

%

\section{Strategic Ability in Stochastic MAS}\label{sec:patl}

We start by recalling the basic definitions of stochastic multi-agent models and strategic play~\cite{Boutilier99mmdp,Chen07PATL,Huang12probabilisticATL}.  In our presentation, we follow mainly~\cite{Belardinelli24PATL}.

\subsection{Probabilistic Models for MAS}\label{sec:probMAS}

Fix finite non-empty sets $\Agt$ of agents $a, a',\ldots$; $\Act$ of actions $\alpha, \alpha',\ldots$; and
$\Props$ of atomic propositions $p, p',\ldots$.
We  write $\profile{o}$ for a tuple $(o_{a})_{{a}\in\Agt}$ of objects, one for each agent; such tuples are called \emph{profiles}. A \emph{joint action} or \emph{move} $\mov$ is an element of $\Act^{\Agt}$.
Given a profile $\profile{o}$ and $\coalition\subseteq\Agt$, we let $o_\coalition$ be the components for the agents in  $\coalition$.
Moreover, we use $\Agt_{-\coalition}$ as a shorthand for $\Agt\setminus\coalition$. 

\smallskip
\para{Distributions.} Let $X$ be a finite non-empty set. A \emph{(probability) distribution} over $X$ is a function $\distribution:X \to [0,1]$ such that $\sum_{x \in X} \distribution(x) = 1$. $\Dist(X)$ is the set of distributions over $X$. We write $x \in \distribution$ for $\distribution(x) > 0$.
If $\distribution(x) = 1$ for some element $x \in X$, then $\distribution$ is a \emph{point (a.k.a. Dirac) distribution}.
 If $\distribution_i$ is a distribution over $X_i$, then, writing $X = \prod_{i} X_i$, the \emph{product distribution} of the $\distribution_i$ is the distribution $\distribution:X \to [0,1]$ defined by $\distribution(x) = \prod_{i} \distribution_i(x_i)$.

\para{Markov Chains. }
  A \emph{Markov chain} $M$ is a tuple $(\States,\distribution)$ where $\States$ is a set of states and $\distribution \in \Dist(\States \times \States)$ is a distribution. The values $\distribution(s,t)$ are called \emph{transition probabilities} of $M$.

\para{Stochastic Concurrent Game Structures. }
  A \emph{stochastic concurrent game structure with imperfect information} (or simply \emph{iCGS})
  $\System$ is a tuple 
  $(
  \States, \legal, 
  \trans, \val, \{\obsrel\}_{{a}\in\Agt})$ where
  (i) $\States$ is a finite, non-empty set of \emph{states};
   (ii) $\legal: \States \times \Agt \to 2^\Act\setminus\{\emptyset\}$ is a function defining the available actions for each agent in each state, i.e., the repertoires of choices. We write $\legal(\state)$ for the set of tuples $(\legal(\state, {a}))_{{a}\in\Agt}$. It is usually assumed that $\legal(\state,{a}) = \legal(\state',{a})$ whenever $\state \obsrel[a] \state'$ (see below);
  (iii)
  for each state $\state \in \States$ and each  move $\mov \in \legal(\state)$, the \emph{stochastic transition function} $\trans$ gives the (conditional) probability $\trans(\state, \mov)$ of a transition from state $\state$ for all $\state' \in \States$ if each player ${a} \in \Agt$ plays the action $\mov_a$; we also write this probability as $\trans(\state, \mov)(\state')$ to emphasize that $\trans(\state, \mov)$ is a probability distribution on $\States$;
  (iv) $\val:\States \to 2^{\Props}$ is a \emph{labelling function};
  (v)
    $\obsrel\;\subseteq \States\times\States$ is an equivalence relation called the  \emph{observation relation} of agent ${a}$.

A pointed \CGS is a pair $(\System,\state)$ where \mbox{$\state \in \States$} is a special state designed as initial.
Throughout this paper, we assume that iCGSs are {\em uniform}, that is, if two states are indistinguishable for an agent ${a}$, then ${a}$ has the same available actions in both states. Formally, if $\state \obsrel \state'$ then $\legal(\state, {a}) = \legal(\state', {a})$, for any $\state, \state' \in \States$ and ${a} \in \Agt$. For each state $\state \in \States$ and joint action $\mov \in \legal(\state)$, we also assume 
  that there is a state $\state'\in\States$ such that $\trans(\state, \mov)(\state')$ is non-zero, that is, every state has a successor state from a legal move.

Finally, we say that $\System$ is \emph{deterministic} (instead of stochastic) if every $\trans(\state,\mov)$ is a point distribution.

\para{Plays. }
A \emph{play} in a iCGS $\System$ is an infinite sequence $\path=\state_0 \state_1 \cdots$ of states
such that there exists a sequence $\mov_0 \mov_1 \cdots$ of joint-actions such that for every $i \geq 0$, $\mov_i \in \legal(\state_{i})$ and  $\state_{i+1} \in \trans(\state_i,\mov_i)$ (\ie, $\trans(\state_i,\mov_i)(\state_{i+1} )>0$).
We write $\path_i$ for state $\state_i$,
$\path_{\geq i}$ for the suffix of
$\path$ starting at position $i$.
Finite prefixes of plays are called \emph{histories}, and the set of all histories is denoted $\History$. Write $\last(\history)$ for the last state of a history $\history$.

\para{Strategies. }
A (general) \emph{probabilistic strategy} for agent ${a}\in\Agt$ is a  function $\str_{a}:\History \to  \Dist(\Act)$ that maps each history to a probability distribution over the agent's actions. It is required that $\str_{a}(\history)(\act) = 0$ if $\act \not \in \legal(\last(\history),{a})$.
We denote the set of ${a}$'s general strategies by $\setstrat_{a}$.

A \emph{memoryless uniform  probabilistic strategy} for an agent ${a}$ is a function $\sigma_{a}: \States \to \Dist(\Act)$, in which:
(i) for each $\state$, we have $\str_{a}(\state)(\act) = 0$ if $\act \not \in \legal(\state,{a})$; and
(ii) for all positions $\state,\state'$ such that $\state\obsrel\state'$, we have $\str_{a}(\state)=\str_{a}(\state')$. We let $\setstrat_a$ be the set of memoryless uniform strategies for agent ${a}$.
We call a memoryless strategy $\str_{a}$ \emph{deterministic} if $\str_{a}(\state)$ is a point distribution for every $\state$.

A \emph{collective strategy} for agents $A\subseteq\Agt$ is a tuple of strategies $\str_{a}$, one per agent ${a}\in A$.
We denote the set of $A$'s collective general strategies and memoryless uniform strategies, respectively, by $\setstrat_A$ and $\setstrat^r_A$.
Moreover, a \emph{strategy profile} is a tuple $\profile\str = \str_\Agt$ of strategies for all the agents. We write $\profile\str_{a}$ for the strategy of  ${a}$ in profile $\profile\str$.

\subsection{Probabilistic \ATL and \ATLs}
\label{sec:patl-logic}
Now we present the syntax and semantics of the Probabilistic Alternating-time Temporal Logics \PATLs and \PATL~\cite{Chen07PATL,Huang12probabilisticATL,Belardinelli23PATL,Belardinelli24PATL}, interpreted under the assumption of imperfect information. Again, we follow~\cite{Belardinelli24PATL} in our presentation. Note that~\cite{Belardinelli24PATL} adopts the {\em objective} semantics of strategic ability, where the coalition is supposed to have a strategy that works from the initial state of the game. In contrast, \cite{Huang12probabilisticATL}~uses the {\em subjective} semantics of strategic ability, where the agents need a strategy that wins from the all the observationally equivalent states.\footnote{
  For a more thorough discussion of objective vs.~subjective ability, cf.~\cite{Bulling14comparing-jaamas,Agotnes15handbook}. }
In this paper, we consider both accounts, as they are equally relevant in the literature. In particular, we integrate the {\em objective} and {\em subjective} semantics of probabilistic ability into a single framework.

\begin{definition}[\PATLs]\label{def:ATLsF-syntax}
State formulas $\varphi$
and path formulas $\psi$
are defined by the following grammar, where $p \in \Props$, $\coalition \subseteq \Agt$, $d$ is a rational constant in
$[0, 1]$, and $\bowtie \in
\{\leq, <, >, \geq\}$:
\begin{eqnarray*}
	\varphi  & ::= & p \mid \neg \varphi \mid  {\varphi \lor  \varphi} \mid \coop{\coalition}^{\bowtie d} \psi\\
 	\psi  & ::= & \varphi \mid \neg \psi \mid  {\psi \lor  \psi}  \mid \Next \psi \mid \psi \Until \psi \mid \psi \Release \psi
\end{eqnarray*}
Formulas in \PATLs are all and only the state formulas $\varphi$.
\end{definition}


An important syntactic restriction of \PATLs, namely  \PATL, is obtained by restricting path formulas as follows:
\begin{eqnarray*}
 	\psi  & ::= & \Next \varphi \mid \varphi \Until \varphi \mid \varphi \Release \varphi
\end{eqnarray*}
%
which is tantamount to the following grammar for state formulas:
	\begin{align*}   	
		\varphi ::= p \mid \neg \varphi  \mid  \varphi \lor \varphi \mid \coop{\coalition}^{\bowtie d} \Next \varphi \mid \coop{\coalition}^{\bowtie d}(\varphi \Until \varphi)
  \mid \coop{\coalition}^{\bowtie d}(\varphi \Release \varphi)
  \end{align*}
where again $p \in \Props$, $\coalition \subseteq \Agt$, and $\bowtie \in \{\leq, <, >, \geq\}$.

Formulas of \PATL and \PATLs are interpreted over iCGSs. 

\para{Probability Space on Outcomes. }
An \emph{outcome} of a strategy $\str_A$ and a state $\state$
is a set of probability distributions over infinite paths, defined as follows.

First, by an \emph{outcome path of a strategy profile $\profile\str$ and state $\state$}, we refer to every play $\path$ that starts with $\state$ and is extended by letting each agent follow their strategies in $\profile\str$, i.e.,
$\path_{0} = \state$, and for every $k \geq 0$ there exists $\mov_k \in
\profile\str(\path_k)$ such that $\path_{k+1} \in \trans(\path_k,\mov_k)$.
The set of {outcome paths} of strategy profile $\profile\str$ and state $\state$ is denoted as $outpaths(\profile\str,\state)$.
A given iCGS $\System$, strategy profile $\profile{\str}$, and state $\state$ induce an infinite-state Markov chain
$M_{\profile{\str},\state}$ whose states are the finite prefixes of plays in $outpaths(\profile{\str},\state)$.
Such finite prefixes of plays are actually \emph{histories}.
Transition probabilities in  $M_{\profile{\str},\state}$ are defined as  $p(\history,\history\state')=\sum_{\mov\in\Act^\Agt}
\profile{\str}(\history)(\mov) \cdot \trans(\last(\history),\mov)(\state')$.
The Markov chain $M_{\profile{\str},\state}$ induces a canonical probability space on
its set of infinite paths~\cite{Kemeny76stochastic}, and thus also on $outpaths(\profile{\str},\state)$.~\footnote{
  This is a classical construction, see for instance~\cite{Clarke18principles,Berthon20alternating}. } 

Given a coalitional strategy $\str_{\coalition} \in \prod_{{a} \in \coalition} \setstrat_a$, we
define its {\em objective outcome} from state $\state \in \States$ as the set
$out_{o,\coalition}(\str_{\coalition},\state) =
  \{out((\str_{\coalition},\str_{\Agt\setminus\coalition}),\state) \mid \str_{\Agt\setminus\coalition} \in \setstrat_{\Agt\setminus\coalition} \}$
of probability measures consistent with strategy $\str_{\coalition}$ of the players in $\coalition$.
Note that the opponents can use any general strategy for $\str_{\Agt\setminus\coalition}$, even if $\coalition$ must employ only uniform memoryless strategies for $\str_{\coalition}$.

The {\em subjective outcomes} are then defined as the set
\begin{equation}
out_{s,\coalition}(\str_{\coalition},\state) = \bigcup_{\state' \sim_a \state, a \in \coalition} out_{o, \coalition}(\str_\coalition,\state') \label{subj}
\end{equation}

We will use $\mu^{\str_{\coalition}}_{x,\state}$ to range over the elements of $out_{x,\coalition}(\str_{\coalition},\state)$, for $x \in \{s, o\}$.


\para{Semantics.}
For $x$ equal to either $s$ or $o$, state and path formulas in \PATLs are interpreted in a iCGS $\System$
  and a state $\state$, resp.~path $\path$, according to the $x$-interpretation of strategy operators, as follows (clauses for Boolean connectives are omitted as immediate):
\begingroup
\allowdisplaybreaks
\begin{align*}
 \System,\state &\models_x p & \text{ iff } & p \in \val(\state)\\
\System, \state  & \models_x \coop{\coalition}^{\bowtie d} \psi & \text{ iff }  & 
\exists \profile{\str_{\!\coalition}} \in \prod_{{a} \in\coalition} \setstrat_a  \text{ such that }
\\ & & & \forall \mu^{\profile{\str_{\!\coalition}}}_{x,\state} \in out_{x,\coalition}(\profile{\str_{\!\coalition}},\state)
\text{, }
\\ & & &
\mu^{\profile{\str_{\!\coalition}}}_{x,s}(\{\path \mid \System,\path \models_x \psi\}) \bowtie d \\
 \System,\path &\models_x \Next \psi & \text{ iff } & \System,\path_{\geq 1} \models_x \psi
 \\
\System, \path  & \models_x \psi_1 \Until \psi_2 & \text{ iff } &  \exists k \geq 0 \text{ s.t. } \System,\path_{\geq k} \models_x \psi_2 \text{ and }
\\ & & &
\forall j \in [0,k)\, \,  \System,\path_{\geq j}\models_x \psi_1\\
\System, \path  & \models_x \psi_1 \Release \psi_2 & \text{ iff } &  \forall k \geq 0, \System,\path_{\geq k} \models_x \psi_2 \text{ or }
\\ & & &
\exists j \in [0,k) \text{ s.t. }  \System,\path_{\geq j}\models_x \psi_1
\end{align*}
\endgroup

%
%

\begin{remark}
Note that standard \ATLs formulas $\coop{\coalition}\psi$ can be interpreted in stochastic iCGS by simply ignoring the probabilities of transitions, i.e., via a projection of the probabilistic transition relation $\trans$ to the non-probabilistic relation $T$ defined by $(\state,\mov,\state') \in T$ iff $\trans(\state, \mov)(\state') > 0$.
\end{remark}

\begin{remark}
Moreover, the logic of Probabilistic \CTLs (\PCTLs) can be embedded in \PATLs by assuming $\Probpath^{\bowtie d} \psi \equiv \coop{\emptyset}^{\bowtie d} \psi$.
\end{remark}

\section{Towards Assumptions and Guarantees for Probabilistic Success}\label{sec:towards-idea}


Now we can turn to assume-guarantee reasoning for probabilistic abilities, expressed in \PATLs.
In this work, we focus on formulas of type $\coop{\coalition}^{\ge \prob}\varphi$ and $\coop{\coalition}^{> \prob}\varphi$.
That is, we will construct our schemes to address the verification of formulas that require the coalition to succeed with probability at least (resp.~better than) a given threshold $\prob$.
Arguably, this kind of requirements is most natural, and most commonly used when characterizing and/or verifying agents' play in a stochastic environment.

We also consider only \emph{deterministic strategies of the coalition}. Extending the scheme to reasoning about probabilistic strategies is nontrivial, and planned for future work. Note that, in our case, it suffices to consider only \emph{deterministic} responses of the opponents, cf.~\cite{Belardinelli24PATL} for the formal justification.

\subsection{Idea}

The main idea is to extend the schemes presented in Section~\ref{sec:assumptions} from reasoning about strategies that \emph{surely win} to strategies that \emph{win with probability of at least $\prob$}. That is, instead of inferring that the coalition has a strategy with all outcome paths satisfying $\psi$, we want the scheme to conclude that the set of outcome paths satisfying $\psi$ has the probability measure of at least $\prob$ (resp. more than $\prob$).

For a path formula $\psi$, let $\denotation{\psi}_\state$ denote the \emph{extension} of $\psi$, i.e., the set of paths starting from $q$ and satisfying $\psi$.
We also assume that $\coalition=\set{1,\dots,m}$ without loss of generality.
Notice that the assume-guarantee schemes of~\cite{Mikulski22AGV} were based on the following reasoning idea:
\begin{enumerate}
\item Prove that, for each member $i$ of $\coalition$ (resp.~each relevant subset of $\coalition$), there is a strategy $\str_i$ such that $\out(\state,\str_i) \subseteq \denotation{\psi_i}_\state$, assuming that the other modules behave according to the assumption $A_i$.

\item Prove, for each $i\in\coalition$, that the behaviour of the other modules is indeed correctly approximated by $A_i$.

\item The above strategies $\str_i$ can be collected into a coalitional strategy $\str_\coalition$, and for each $i\in\coalition$ we obviously have that $\out(\state,\str_\coalition) \subseteq \out(\state,\str_i) \subseteq \denotation{\psi_i}_\state$.
    Thus, $\out(\state,\str_\coalition) \subseteq \bigcap_{i\in\coalition} \denotation{\psi_i}_\state$.

\item In consequence, we can conclude that $\coop{\coalition}\psi$ for $\psi \equiv \bigwedge_{i\in\coalition} \psi_i$.
\end{enumerate}

Ideally, we want to lift the above reasoning to the case where we conclude that $\coalition$ enforces $\psi$ \emph{with probability} of at least (resp. more than) $\prob$.
More formally, instead of inferring that $\out(\state,\str_\coalition) \subseteq \denotation{\psi}_\state$, we need to prove that the probability measure of $\out(\state,\str_\coalition,\str_{\Agt\setminus\coalition}) \cap \denotation{\psi}_\state$  is at least (resp. more than) $\prob$ for every counterstrategy $\str_{\Agt\setminus\coalition}$ of the agents outside $\coalition$.
Unfortunately, a straightforward extension does not work.

\subsection{Combining Outcome Sets While Preserving Probabilities}

Consider a simple extension of the previous approach, along the following idea:
\begin{enumerate}
\item\label{it:agv-indiv-probs} Prove that, for each $i\in\coalition$ (resp.~each relevant subset of $\coalition$), there is a strategy $\str_i$ that achieves $\psi_i$ with probability at least $\prob_i$ (i.e., such that $\mu(\out(\state,\str_i) \cap \denotation{\psi_i}_\state) \ge \prob_i$), assuming that the other modules behave according to the assumption $A_i$.

\item Prove that the behaviour of the other modules is correctly approximated by $A_i$.

\item The above strategies $\str_i$ can be collected into a coalitional strategy $\str_\coalition$ such that, for every $\str_{\Agt\setminus\coalition}$, we have $\mu(\out(\state,\str_\coalition,\str_{\Agt\setminus\coalition}) \cap \bigcap_{i\in\coalition} \denotation{\psi_i}_\state) \ge f(\prob_1,\dots,\prob_m)$, where $f$ is a function of the probabilities obtained in step~(\ref{it:agv-indiv-probs}).

\item In consequence, we can conclude that $\coop{\coalition}^{\ge \prob}\psi$ for $\prob=f(\prob_1,\dots,\prob_m)$ and $\psi \equiv \bigwedge_{i\in\coalition} \psi_i$.
\end{enumerate}
The scheme for $\coop{\coalition}^{>\prob}\psi$ would be constructed analogously.

\begin{figure}[t]\centering
\includegraphics[scale=0.25]{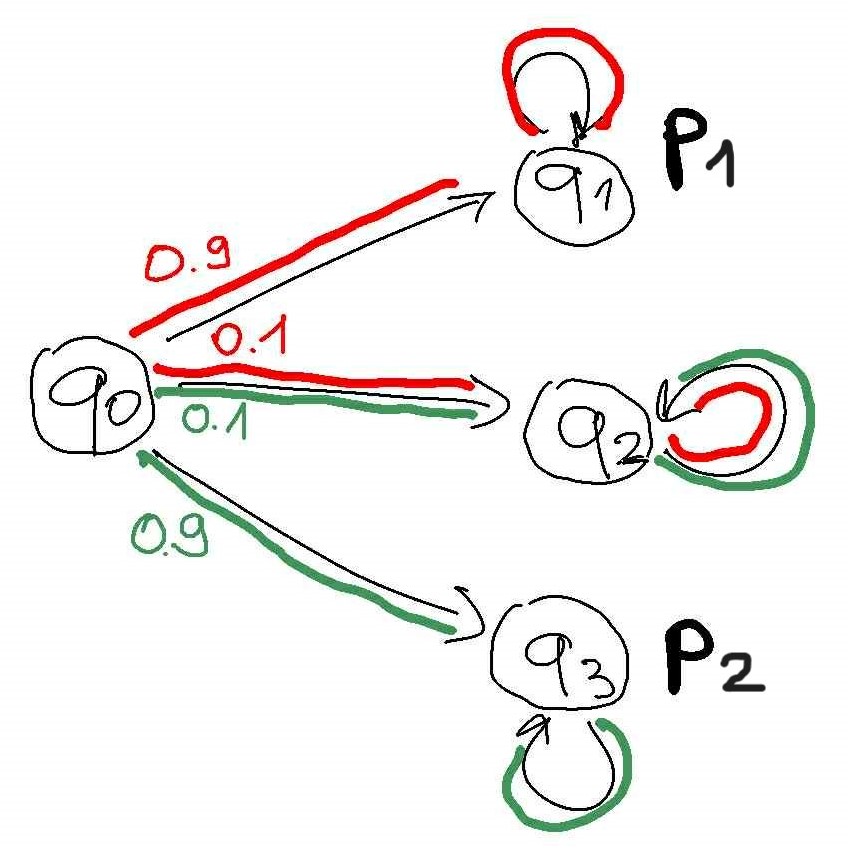}
\caption{Probabilities do not combine for intersection of outcome sets: case 1}
\label{fig:out-intersection-1}
\end{figure}

Unfortunately, the above reasoning is not sound. The problem is that, even is the outcome sets $\out_1,\dots,\out_2$ guarantee $\psi_1,\dots,\psi_m$ with probabilities $\prob_1,\dots,\prob_m$, their combination might consist of paths where not all of $\psi_1,\dots,\psi_m$ holds, cf.~the counterexample in Figure~\ref{fig:out-intersection-1}.
The model includes two agents $\Agt=\set{1,2}$.
The red lines depict the probability distribution for the best strategy of agent $1$ (against the most damaging response of agent $2$) to obtain $\Sometm\prop{p_1}$.
Similarly, the green lines show the probability distribution for the best strategy of agent $2$ (against the most damaging response of agent $1$) to obtain $\Sometm\prop{p_2}$.
Clearly, $\model,q_0 \models \coop{1}^{\ge 0.9}\Sometm\prop{p_1}$ and $\model,q_0 \models \coop{2}^{\ge 0.9}\Sometm\prop{p_2}$.
At the same time $\model,q_0 \not\models \coop{1,2}^{\ge\rho}(\Sometm\prop{p_1} \land \Sometm\prop{p_2})$ for all $\rho\neq 0$, because no path in $\model$ satisfies both $\Sometm\prop{p_1}$ and $\Sometm\prop{p_2}$.

\subsection{Solution (Semantic Formulation)}\label{sec:solution-semantic}

As a solution, we propose to split the assume-guarantee scheme into reasoning about (non-probabilistic) outcome sets of $\coalition$'s strategies, and probabilistic reasoning about the combination (i.e., intersection) of those outcome sets:
\begin{enumerate}
\item Prove that, for each member $i$ of $\coalition$ (resp.~each relevant subset of $\coalition$), there is a strategy $\str_i$ such that $\out(\state,\str_i) \subseteq \denotation{\psi_i}_\state$, assuming that the other modules behave according to the assumption $A_i$.

\item Prove that the behaviour of the other modules is correctly approximated by $A_i$.

\item\label{it:intersection} The above strategies $\str_i$ can be collected into a coalitional strategy $\str_\coalition$ such that $\out(\state,\str_\coalition) \subseteq \bigcap_{i\in\coalition} \denotation{\psi_i}_\state$.

\item\label{it:pctl} Prove that the probability of getting $\psi$ within the paths satisfying $\bigwedge_{i\in\coalition} \psi_i$ must be at least $\prob$.
    Formally, this amounts to proving that the conditional probability $\mu(\denotation{\psi} \mid \bigcap_{i\in\coalition} \denotation{\psi_i}) \ge \prob$ for all possible choices of the agents.

\item In consequence, we can conclude that $\coop{\coalition}^{\ge \prob}\psi$.
\end{enumerate}
The scheme for $\coop{\coalition}^{>\prob}\psi$ is constructed analogously.
We formalize the idea in Section~\ref{sec:agv-patl}.

%

\section{Assume-Guarantee Verification for Probabilistic ATL}\label{sec:agv-patl}

Now we propose our new assume-guarantee schemes for \PATLs, adapting the ones in~\cite{Mikulski22AGV} to verification of probabilistic properties.

\subsection{Assume-Guarantee Reasoning for \PATLs: from Individual to Coalitional Abilities}\label{sec:agv-prob-individual}


Let $\psi_{i}$ be an \LTL formula (without ``next''), where atomic propositions are local valuations of variables in $M\ag{i}$.
Also, let $x\in\{\ir,\iR\}$.
The scheme is formalized through a sequence of rules $\bf{PR_k}$ which rely on the behaviour of the neighbourhoods of coalition $\coalition$, limited by ``distance'' $k$:

\[
\bf{PR_k}\;\;\;
\begin{array}{c}
\forall_{i\in \coalition}\; (M\ag{i} | A_i, R\ag{i}) \,\satisf[x]\, \coop{i} \psi_{i}\\
\forall_{i\in \coalition}\; \compos\ag{i}_k \,\models\, A_i\\
(M\ag{1} | ... | M\ag{n},R\ag{1},\dots,R\ag{n}) \,\satisf[x]\, \Probpath^{\ge \prob_1} (\psi \land \bigwedge_{i\in \coalition}\psi_{i})\\
(M\ag{1} | ... | M\ag{n},R\ag{1},\dots,R\ag{n}) \,\satisf[x]\, \Probpath^{\le \prob_2} (\bigwedge_{i\in \coalition}\psi_{i})
\smallskip \\
\hline
(M\ag{1} | ... | M\ag{n},R\ag{1},\dots,R\ag{n}) \,\satisf[x]\, \coop{\coalition}^{\ge \prob_1/\prob_2} \psi
\end{array}
\]

The following theorem says that, if each coalition member together with its assumption satisfies the decomposition of the formula, and its neighborhood satisfies the assumption, then the original verification task must return ``true.''

\begin{theorem}\label{t:prob-rksound}
The rule $\bf{PR_k}$ is sound.
\end{theorem}
\begin{proof}
Analogous to~\cite[Theorem~1]{Mikulski22AGV}.
\end{proof}

Similarly to its non-probabilistic analogue, there does not always exist $k<n$ for which the rule $\bf{PR_k}$ is complete,
even in a very weak sense, where we only postulate the \emph{existence} of appropriate assumptions.

\begin{theorem}\label{t:prob-rkincomplete}
The scheme consisting of rules $\set{\bf{PR_k} \mid k\in\Nat}$ is in general not complete.
\end{theorem}
\begin{proof}
Analogous to~\cite[Theorem~2]{Mikulski22AGV}.
\end{proof}

There are two main challenges in applying the above scheme.
First, one must define the right assumptions and decompose the verified formula in a good way.
Secondly, the verification of $\Probpath^{\ge \prob} (\psi \land \bigwedge_{i\in \coalition}\psi_{i})$ is \emph{with respect to the whole model}, which is what we wanted to avoid in the first place.
However, this part of the rule concerns the verification of \PCTL or \PCTLs formulas, which is a significantly cheaper problem than for \PATL or \PATLs\extended{~\cite{Chen07PATL}}. Moreover, efficient assume-guarantee reasoning schemes already exist for Probabilistic \CTL~\cite{Kwiatkowska10assGuar,Komuravelli12AGV+AR-prob,Chatterjee14CEGAR+AGV-prob}.

Still, it would be much better if we could avoid generating the whole global model of the system.
We show how this can be partially achieved in the next subsection.

\subsection{Replacing Strategic Model Checking with Strategy Synthesis}\label{sec:agv-prob-synthesis}

An important observation is that all known model checking algorithms for \ATL can be also used for strategy synthesis. In order to verify if $\coalition$ has a strategy to enforce $\psi$, the standard approach is to attempt the synthesis of a winning strategy and return yes iff the attempt has been successful\extended{~\cite{XXXXXXX}}. Alternatively, one can try to verify $\coop{\coalition}\psi$ by some kind of approximation\extended{~\cite{FIXPAPPR,ABSTRACTION}}, but also in those cases it is possible to extract a winning strategy from the approximation output.
In consequence, we can use the strategy being the witness of $(M\ag{i} | A_i, R\ag{i}) \,\satisf[x]\, \coop{i} \psi_{i}$ to simplify the \PCTL verification step\linebreak $(M\ag{1} | ... | M\ag{n},R\ag{1},\dots,R\ag{n}) \,\satisf[x]\, \Probpath^{\ge \prob} (\psi \land \bigwedge_{i\in \coalition}\psi_{i})$ in the assume-guarantee rule presented above.

To do this, we introduce additional notation:
\begin{center}
$\denotation{M,\state \satisf[x] \coop{\coalition} \psi} = $ any $x$-strategy $\str_\coalition$ enforcing $\psi$ from $\state$ (if it exists), otherwise $\bot$.
\end{center}
We also observe that, if $\denotation{(M\ag{i} | A_i, R\ag{i}) \,\satisf[x]\, \coop{i} \psi_{i}} = \str_i$, then the model\linebreak
$(\str_1 | \dots | \str_m | M\ag{m+1} | \dots | M\ag{n},R\ag{m+1},\dots,R\ag{n})$ satisfies the following properties:
\[
\begin{array}{c}
\mu(\denotation{\bigwedge_{i\in \coalition}\psi_{i}}) = 1 \\
\mu(\denotation{\psi \land \bigwedge_{i\in \coalition}\psi_{i}}) = \mu(\denotation{\psi})
\end{array}
\]

Thus, the assume-guarantee rule of Section~\ref{sec:agv-prob-individual} can be updated as follows:
\[
\bf{PRsynt_k}\;\;\;
\begin{array}{c}
\forall_{i\in \coalition}\; \denotation{(M\ag{i} | A_i, R\ag{i}) \,\satisf[x]\, \coop{i} \psi_{i}} = \str_i \neq \bot \\
\forall_{i\in \coalition}\; \compos\ag{i}_k \,\models\, A_i\\
(\str_1 | \dots | \str_m | M\ag{m+1} | \dots | M\ag{n},R\ag{m+1},\dots,R\ag{n}) \,\satisf[x]\, \Probpath^{\ge \prob} \psi\\
\hline
(M\ag{1} | \dots | M\ag{n},R\ag{1},\dots,R\ag{n}) \,\satisf[x]\, \coop{\coalition}^{\ge \prob} \psi
\end{array}
\]

\begin{theorem}\label{t:prob-rkwithsynthesis}
The rule $\bf{PRsynt_k}$ is sound, but in general not complete.
\end{theorem}
\begin{proof}
Straightforward extension of the proofs for Theorems~\ref{t:prob-rksound} and~\ref{t:prob-rkincomplete}.
\end{proof}

\subsection{Sound and Complete Assume-Guarantee Reasoning for \PATLs}\label{sec:agv-prob-compl}

Again, let $\model=(M\ag{1},R\ag{1}),\ \dots,\ (M\ag{n},R\ag{n})$ be a system composed of asynchronous agents,
and $\{P_1, \ldots, P_k : P_i\subseteq\{1,2,\ldots,n\}\}$ be a partitioning of coalition $\coalition$.
%
By $\model\ag{P_i}$ we denote the system composed of all the agents in $P_i = \{i_1,\ldots,i_s\}$, i.e., $(M\ag{P_i}=M\ag{i_1}|\ldots|M\ag{i_s}, R\ag{i_1},\ldots,R\ag{i_s})$.
%
We recall the notion of neighbourhood sets from Section~\ref{sec:atl+agv}:\
\begin{itemize}
\item $N^{P_i}_1=(\bigcup_{i\in P_i}N\ag{i}_1)\setminus P_i$,\
$N^{P_i}_k=(N^{P_i}_{k-1}\cup\bigcup_{j\in N^{P_i}_{k-1}}N\ag{j}_1) \setminus P_i$ for $k>1$,\
\item $\compos^{P_i}_k=M\ag{x_1} | ... | M\ag{x_s}$ for $N^{P_i}_k=\{x_1,...,x_s\}$.
\end{itemize}

Let $x\in \{\ir,\iR\}$. The generalized assume-guarantee rule is defined as follows:

\[
\bf{PPart^P_k}\;\;\;
\begin{array}{c}
\forall_{P_i\in P}\; \denotation{(M\ag{P_i} | A_i,R\ag{i_1},\ldots,R\ag{i_s}) \,\models_{x}\, \coop{P_i} \bigwedge_{j\in P_i}\psi_j} = \str_{P_i} \neq \bot \\
\forall_{P_i\in P}\; \compos^{P_i}_k \,\models\, A_i\\
(\str_1 | \dots | \str_m | M\ag{m+1} | \dots | M\ag{n},R\ag{m+1},\dots,R\ag{n}) \,\satisf[x]\, \Probpath^{\ge \prob} \psi\\
\hline
(M\ag{1} | ... | M\ag{n},R\ag{1},\dots,R\ag{n}) \,\satisf[x]\, \coop{\coalition}^{\ge \prob} \psi
\end{array}
\]


\begin{theorem}[\cite{Mikulski22AGV}]\label{t:partsound}
The rule $\bf{PPart^P_k}$ is sound.
\end{theorem}
\begin{proof}
Analogous to~\cite[Theorem~3]{Mikulski22AGV}.
\end{proof}

\begin{theorem}[\cite{Mikulski22AGV}]\label{t:complete}
There exist a partition set $P$ and $k\leq n$ such that the rule $\bf{PPart^P_k}$ is complete.
\end{theorem}
\begin{proof}
Analogous to~\cite[Theorem~4]{Mikulski22AGV}.
\end{proof}

\subsection{Verification of Nested Strategic Operators}\label{sec:nested}

Similarly to~\cite{Mikulski22AGV}, we point out that the schemes $\bf{PR_k}$ and $\bf{PPart^P_k}$ can be extended to the whole language of \ATLsX through the standard recursive model checking algorithm that verifies subformulas bottom-up.
Such recursive application of the method to the verification of $\model \models \phi$ proceeds as follows:
\begin{itemize}
\item For each strategic subformula $\phi_j$ of $\phi$, do assume-guarantee verification of $\phi_j$ in $\model$, and label the states where $\phi_j$ holds by a fresh atomic proposition $\prop{p_j}$;
\item Replace all occurrences of $\phi_j$ in $\phi$ by $\prop{p_j}$, and do assume-guarantee verification of the resulting formula in $\model$.
\end{itemize}

The resulting algorithm is sound (and complete in case of scheme \textbf{PPart}), though there is the usual price to pay in terms of computational complexity.

\section{The Uses of Incompetence}\label{sec:agv-patl-onlyachieving}

We have just shown that reasoning about \emph{upper approximations} of outcomes for strategies $\str_1,\dots,\str_m$ allows one to draw valid conclusions about probabilities of temporal patterns in the combined coalitional strategy.
In this section, we propose an alternative scheme based on reasoning about \emph{lower approximations} of the outcome sets.

\subsection{Semantic Formulation}\label{sec:lower-semantic}

In order to achieve this, we need individual lower approximations to combine into coalitional lower approximations of outcomes.
That is, individual lower approximations must produce (via intersection) a lower approximation of the coalitional outcome for $\str_\coalition = (\str_1,\dots,\str_m)$. It is not the case for outcome \emph{plays}, i.e., we only have that $\out(\state,\str_\coalition) \subseteq \bigcap_{i\in\coalition} \out(\state,\str_i)$, but not vice versa. However, we do have the property for sets of outcome \emph{traces}:

\begin{proposition}
If $\str_\coalition = (\str_1,\dots,\str_m)$, then
$\outtr(\state,\str_\coalition) = \bigcap_{i\in\coalition} \outtr(\state,\str_i)$.
\end{proposition}

Throughout this section, we assume that the extension $\denotation{\psi}_\state$ of formula $\psi$ contains traces rather than paths satisfying $\psi$.
This leads to the following semantic reformulation of the scheme:
\begin{enumerate}
\item\label{it:ability-reverse} Prove that, for each member $i$ of $\coalition$ (resp.~each relevant subset of $\coalition$), there is a strategy $\str_i$ such that \superemph{$\denotation{\psi_i}_\state \subseteq \outtr(\state,\str_i)$}, assuming that the other modules behave according to the assumption $A_i$.

\item Prove that the behaviour of the other modules is correctly approximated by $A_i$.

\item\label{it:pctl-new} Prove that the probability of getting $\psi$ in the set of traces satisfying $\bigwedge_{i\in\coalition} \psi_i$ is at least $p$, i.e., $\mu(\denotation{\psi} \cap \bigcap_{i\in\coalition} \denotation{\psi_i}) \ge p$.

\item The above strategies $\str_i$ can be collected into a coalitional strategy $\str_\coalition$ such that
    \superemph{$\mu( \denotation{\psi} \cap  \outtr(\state,\str_\coalition))
     = \mu( \denotation{\psi} \cap \bigcap_{i\in\coalition} \denotation{\psi_i}_\state)
     \ge p$}.
    In consequence, we can conclude that $\coop{\coalition}^{\ge p}\psi$.
\end{enumerate}
%

Notice that the requirement in the new point~(\ref{it:ability-reverse}) cannot be captured by the \ATLs formula $\coop{i}\psi_i$ anymore. In fact, it cannot be expressed by any formula of \ATLs.\footnote{
  We will show this formally in Section~\ref{sec:incompetence}. }
Intuitively, $\coop{i}\psi_i$ captures the existence of an $i$'s strategy whose outcome is covered by $\phi_i$, whereas in  point~(\ref{it:ability-reverse}) we need to establish the existence of an $i$'s strategy whose outcome covers $\phi_i$. In other words, such that $i$ might possibly enforce $\psi_i$, but certainly does not enforce any \emph{stronger} objective.
We will formalize this kind of properties in Section~\ref{sec:incompetence}.


\subsection{A Logic of Willful Incompetence}\label{sec:incompetence}


We introduce a new variant of alternating-time temporal logic, \onlyATLs, with strategic modalities $\cooponly{\coalition}\psi$, expressing that $\coalition$ has a strategy that only enforces properties weaker or equivalent to $\psi$.

\subsection{Syntax and Semantics}

Formally, we take the syntax of \ATLs and replace the standard modalities $\coop{\coalition}$ by $\cooponly{\coalition}$. Moreover, we define the semantics of $\cooponly{\coalition}$ through the following clause:
\begin{description}
%
\item[{$\model,q \satisf[x] \cooponly{\coalition} \psi$}] if there exists an $x$-strategy $\str_\coalition$ for $\coalition$ such that, for every general counterstrategy $\str_{\Agt\setminus\coalition}$ and any trace $\path \in \traces(\state)$, if $\model,\path \satisf[x] \psi$, then $\path \in \outtr(\state,\str_\coalition,\str_{\Agt\setminus\coalition})$.
%
%
\end{description}
In other words, $\forall_{\str_{\Agt\setminus\coalition}} \dott \denotation{\psi}_\state \subseteq \outtr(\state,\str_\coalition,\str_{\Agt\setminus\coalition})$.
By transposition, we obtain the following alternative definition of $\cooponly{\coalition}$.

\begin{proposition}
\begin{description}
\item[{$\model,q \satisf[x] \cooponly{\coalition} \psi$}] if there exists an $x$-strategy $\str_\coalition$ for $\coalition$ such that\linebreak $\traces(\state) \setminus \bigcap_{\str_{\Agt\setminus\coalition}}\outtr(\state,\str_\coalition,\str_{\Agt\setminus\coalition}) \subseteq \denotation{\neg\psi}_\state$.
\end{description}
\end{proposition}
\begin{proof}
$\model,q \satisf[x] \cooponly{\coalition} \psi$
    \quad\textbf{iff}\quad
$\exists_{\str_\coalition\in\setstrat^x_\coalition}\forall_{\str_{\Agt\setminus\coalition}} \dott
 \denotation{\psi}_\state \subseteq \outtr(\state,\str_\coalition,\str_{\Agt\setminus\coalition})$
    \quad\textbf{iff}\quad
$\exists_{\str_\coalition\in\setstrat^x_\coalition}\forall_{\str_{\Agt\setminus\coalition}} \dott
 \traces(\state) \setminus \outtr(\state,\str_\coalition,\str_{\Agt\setminus\coalition}) \subseteq \traces(\state) \setminus \denotation{\psi}_\state$
    \quad\textbf{iff}\quad
$\exists_{\str_\coalition\in\setstrat^x_\coalition} \dott
 \traces(\state) \setminus \bigcap_{\str_{\Agt\setminus\coalition}}\outtr(\state,\str_\coalition,\str_{\Agt\setminus\coalition}) \subseteq \denotation{\neg\psi}_\state$.
\end{proof}

``Vanilla'' \onlyATL is obtained from \onlyATLs through the usual syntactic restriction.

Arguably, $\cooponly{\coalition}$ has less use in real life than $\coop{\coalition}$. Still, we observe that it can be used, e.g., to express that an employee can hide his/her ability to realize certain tasks in a corporate setting, in order to avoid being assigned those tasks. Similarly, we can express $\coalition$'s ability to keep secret about their monitoring ability in a cybersecurity scenario, etc.
Here, we use the new operator for purely technical reasons, in order to formalize the assume-guarantee scheme outlined there.

Before we move on, we establish relevant metaproperties of \onlyATL.

\subsection{Verification Complexity}

First, we observe that the model checking complexity of \onlyATL is the same as \ATL for the corresponding semantic variants.

\begin{theorem}\label{prop:onlyATL-complexity}
\mbox{}

  \begin{enumerate}
  \item Model checking \onlyATL[\IR] and \onlyATL[\Ir] is \Ptime-complete with respect to the size of the model and the length of the formula.
  \item Model checking \onlyATL[\ir] is \Deltwo-complete with respect to the size of the model and the length of the formula.
  \end{enumerate}
\end{theorem}
\begin{proof}[sketch]
  \begin{enumerate}
  \item Analogous to~\cite{Alur02ATL}; we use a fixpoint algorithm where the pre-image operation checks if all the \emph{non}-outcome transitions end up \emph{outside} of the current solution subset of states.

  \item Analogous to~\cite{Schobbens04ATL,Jamroga06atlir-eumas}. In particular, the upper bound is obtained by recursively guessing the strategy $s_\coalition$ behind $\cooponly{\coalition}\psi$, pruning the transitions \emph{consistent} with the strategy, and checking if $\Apath\neg\psi$ holds in the remaining model.

  \end{enumerate}
\end{proof}

We also conjecture that model checking of \onlyATL[\iR] is undecidable.

\subsection{Expressivity}

Secondly, there are properties expressible in \onlyATL that cannot be expressed in \ATL.
We start by recalling the formal definitions of expressive and distinguishing power.

\begin{definition}[Expressive power and distinguishing power~\cite{Wang09expressive}]
Consider two logical systems ${L}_1$ and ${L}_2$, with their semantics defined
over the same class of models $\mathcal{M}$.
${L}_1$ is \emph{at least as expressive as ${L}_2$} (written ${L}_2 \lexpr {L}_1$) if,
for every formula $\varphi_2$ of ${L}_2$, there exists a formula $\varphi_1$ of ${L}_1$,
such that $\varphi_1$ and $\varphi_2$ are satisfied in the same models from $\mathcal{M}$.

Moreover, ${L}_1$ is \emph{at least as distinguishing as ${L}_2$} (${L}_2 \ldist {L}_1$)
if every pair of models $M,M'\in\mathcal{M}$ that can be
distinguished by a formula of ${L}_2$ can also be distinguished by some formula of ${L}_1$.
\end{definition}
It is easy to see that ${L}_2 \lexpr {L}_1$ implies ${L}_2 \ldist {L}_1$.
By transposition, we also have that ${L}_2 \not\ldist {L}_1$ implies ${L}_2 \not\lexpr {L}_1$.

\begin{theorem}
The distinguishing power of \onlyATL is not covered by \ATL, i.e., $\onlyATL \not\ldist \ATL$.
\end{theorem}
\begin{proof}
Consider two models $M, M'$ with agents $\Agt=\set{1,2}$, states $\States=\set{init,0,1}$, and transitions at state $init$ given by the following transition tables:
\begin{center}
{\large ($M$)}
\qquad
\begin{tabular}{|c|c|c|}
  \hline
  $1\backslash 2$ & $\,\alpha\,$ & $\,\beta\,$
  \\   \hline
  $\alpha$ & $1$ & $1$
  \\   \hline
  $\beta$ & $0$ & $0$
  \\   \hline
  $\gamma$ & $0$ & $0$
  \\   \hline
\end{tabular}
\qquad
\qquad
\qquad
{\large ($M'$)}
\qquad
\begin{tabular}{|c|c|c|}
  \hline
  $1\backslash 2$ & $\,\alpha\,$ & $\,\beta\,$
  \\   \hline
  $\alpha$ & $1$ & $1$
  \\   \hline
  $\beta$ & $0$ & $0$
  \\   \hline
  $\gamma$ & $0$ & $1$
  \\   \hline
\end{tabular}
\end{center}
The only outgoing transitions at states $0,1$ are self-loops.
The sole atomic proposition $\prop{p}$ holds only at state $1$.

Observe that $(M,init)$ and $(M',init)$ satisfy exactly the same formulas of \ATL (and even \ATLs).
On the other hand, $M,init \models \cooponly{1}\Next\prop{p}$ but $M',init \not\models \cooponly{1}\Next\prop{p}$, which concludes the proof.
\end{proof}

The following is a straightforward corollary.

\begin{theorem}
The expressive power of \onlyATL is not covered by \ATL, i.e., $\onlyATL \not\lexpr \ATL$.
\end{theorem}

\subsection{Assume-Guarantee Scheme Based on ``Only Achieving''}

Let $\psi_{i}$ be an \LTL formula (without ``next''), where atomic propositions are local valuations of variables in $M\ag{i}$.
Also, let $x\in\{\ir,\iR\}$.
Again, the scheme is formalized through a sequence of rules $\bf{PRonly_k}$ which rely on the behaviour of the neighbourhoods of coalition $\coalition$, limited by ``distance'' $k$:
\[
\bf{PRonly_k}\;\;\;
\begin{array}{c}
\forall_{i\in \coalition}\; (M\ag{i} | A_i, R\ag{i}) \,\satisf[x]\, \cooponly{i} \psi_{i}\\
\forall_{i\in \coalition}\; \compos\ag{i}_k \,\models\, A_i\\
(M\ag{1} | ... | M\ag{n},R\ag{1},\dots,R\ag{n}) \,\satisf[x]\, \Probpath^{\ge p} (\psi \land \bigwedge_{i\in \coalition}\psi_{i})\\
\hline
(M\ag{1} | ... | M\ag{n},R\ag{1},\dots,R\ag{n}) \,\satisf[x]\, \coop{\coalition}^{\ge p} \psi
\end{array}
\]

The following theorem says that, if each coalition member together with its assumption satisfies the decomposition of the formula, and its neighborhood satisfies the assumption, then the original verification task must return ``true.''

\begin{theorem}\label{t:prob-rk-atlonly}
The rule $\bf{PRonly_k}$ is sound but not complete.
\end{theorem}
\begin{proof}
Analogous to~\cite[Theorem~1 and Theorem~2]{Mikulski22AGV}.
\end{proof}

Similarly to Section~\ref{sec:agv-prob-synthesis}, we can refine the above scheme by using the output of strategy synthesis:
\[
\bf{PRso_k}\;\;\;
\begin{array}{c}
\forall_{i\in \coalition}\; \denotation{(M\ag{i} | A_i, R\ag{i}) \,\satisf[x]\, \cooponly{i} \psi_{i}} = \str_i \\
\forall_{i\in \coalition}\; \compos\ag{i}_k \,\models\, A_i\\
(\str_1 | \dots | \str_m | M\ag{m+1} | \dots | M\ag{n},R\ag{m+1},\dots,R\ag{n}) \,\satisf[x]\, \Probpath^{\ge p} (\psi \land \bigwedge_{i\in \coalition}\psi_{i})\\
\hline
(M\ag{1} | ... | M\ag{n},R\ag{1},\dots,R\ag{n}) \,\satisf[x]\, \coop{\coalition}^{\ge p} \psi
\end{array}
\]

\begin{theorem}\label{t:prob-rk-atlonly}
The rule $\bf{PRso_k}$ is sound but not complete.
\end{theorem}
\begin{proof}
As above.
\end{proof}

\section{Conclusions}\label{sec:conclusions}

In this paper, we present several schemes for assume-guarantee verification of strategic abilities in stochastic multi-agent systems.
The schemes provide means for decomposition of verification for system properties given in Probabilistic \ATLs, typically for agents with imperfect information.
All the schemes are sound for the memoryless as well as perfect recall semantics of abilities under imperfect information.
Moreover, one of them is complete, though in a relatively weak sense.
The schemes have been inspired by~\cite{Lomuscio10assGar,Lomuscio13assGar,Mikulski22AGV}, but the technical formulation is significantly different.
In particular, we have defined a completely novel variant of alternating-time temporal logic, expressing the agents' ability to achieve \emph{no more than $\varphi$}, in order to formalize the decomposition.
Importantly, the properties captured by the new logic cannot be expressed in standard \ATLs, yet at the same time it features the same verification complexity.

The main challenge when using the scheme are: (i) formulation of the right assumptions (i.e., Buchi abstractions of parts of the system), (ii) decomposition of the verified path subformula $\psi$, and (iii) verification of the \PCTLs ``residue'' in steps (3)-(4) of the assume-guarantee inference rules. We observe that the last challenge can be alleviated by using an assume-guarantee scheme for Probabilistic \CTL~\cite{Kwiatkowska10assGuar,Komuravelli12AGV+AR-prob,Chatterjee14CEGAR+AGV-prob}.
We also speculate that the Buchi assumptions from steps (1)-(2) can be used to further simplify the reasoning, but we leave the in-depth study of this idea for future work.


\para{Acknowledgements}
The work has been supported by NCBR Poland and FNR Luxembourg under the PolLux/FNR-CORE project SpaceVote (POLLUX-XI/14/SpaceVote/2023 and C22/IS/17232062/SpaceVote).
For the purpose of open access, and in fulfilment of the grant agreement, the authors have applied CC BY 4.0 license to any Author Accepted Manuscript version arising from this submission.

\bibliographystyle{apalike}
\bibliography{wojtek,wojtek-own}

\end{document}